\documentclass[11pt,twoside]{book} 
\usepackage{asp2008s}
\usepackage{times}
\usepackage{lscape}
\usepackage{epsf}

\usepackage{graphicx}
\usepackage{amssymb}
\usepackage{longtable}
\usepackage[figuresright]{rotating}
\usepackage{multirow}

\newcommand{\simless}{\mathbin{\lower 3pt\hbox {$\rlap{\raise 5pt\hbox{$\char'074$}}\mathchar"7218$}}}

\mainmatter

\newlength{\deftabcolsep}
\begin{document}

\title{NGC\,2362: The Terminus of Star Formation}   %%% Fill in title
\author{S. E. Dahm}   %%% Fill in author names
\affil{Department of Astronomy, California Institute of Technology, MS 105-24, Pasadena, CA 91125, USA}    %%% Fill in author affiliations

\begin{abstract} %%% Abstract to run on from here.
NGC\,2362 is a richly populated Galactic cluster, devoid of natal
molecular gas and dust. The cluster represents the final product of
the star forming process and hosts an unobscured and near-complete
initial mass function. NGC\,2362 is dominated by the O9 Ib multiple
star, $\tau$ CMa, as well as several dozen unevolved B-type
stars. Distributed throughout the cluster are several hundred
suspected intermediate and low-mass pre-main sequence members. Various
post-main sequence evolutionary models have been used to infer an age
of $\sim$5 Myr for the one evolved member, $\tau$ CMa. These estimates
are in close agreement with the ages derived by fitting pre-main
sequence isochrones to the contracting, low-mass stellar population of
the cluster. The extremely narrow sequence of stars, which extends
more than 9 mag in the optical color-magnitude diagram, suggests that
star formation within the cluster occurred rapidly and coevally across
the full mass spectrum. Ground-based near infrared and H$\alpha$
emission surveys of NGC\,2362 concluded that most ($\sim$90\%) of the
low-mass members have already dissipated their optically-thick, inner
($\ll$1 AU) circumstellar disks. {\it Spitzer} IRAC observations of
the cluster have confirmed these results, placing an upper limit on
the primordial, optically thick disk fraction of the cluster at
$\sim$7$\pm$2\%. The presence of circumstellar disks among candidate
members of NGC\,2362 is also strongly mass-dependent, such that no
stars more massive than $\sim$1.2 M$_{\odot}$ exhibit significant
infrared excess shortward of 8 $\mu$m. NGC\,2362 will likely remain a
favored target of ground-based and space-based observations. Its
well-defined upper main sequence, large population of low-mass,
pre-main sequence stars, and the narrow age spread evident in the
color-magnitude diagram ensure its role as a standard model of cluster
as well as stellar evolution.
\end{abstract}

\section{Introduction}

\begin{figure}[h!]
\hspace{-1.0cm}
\plotfiddle{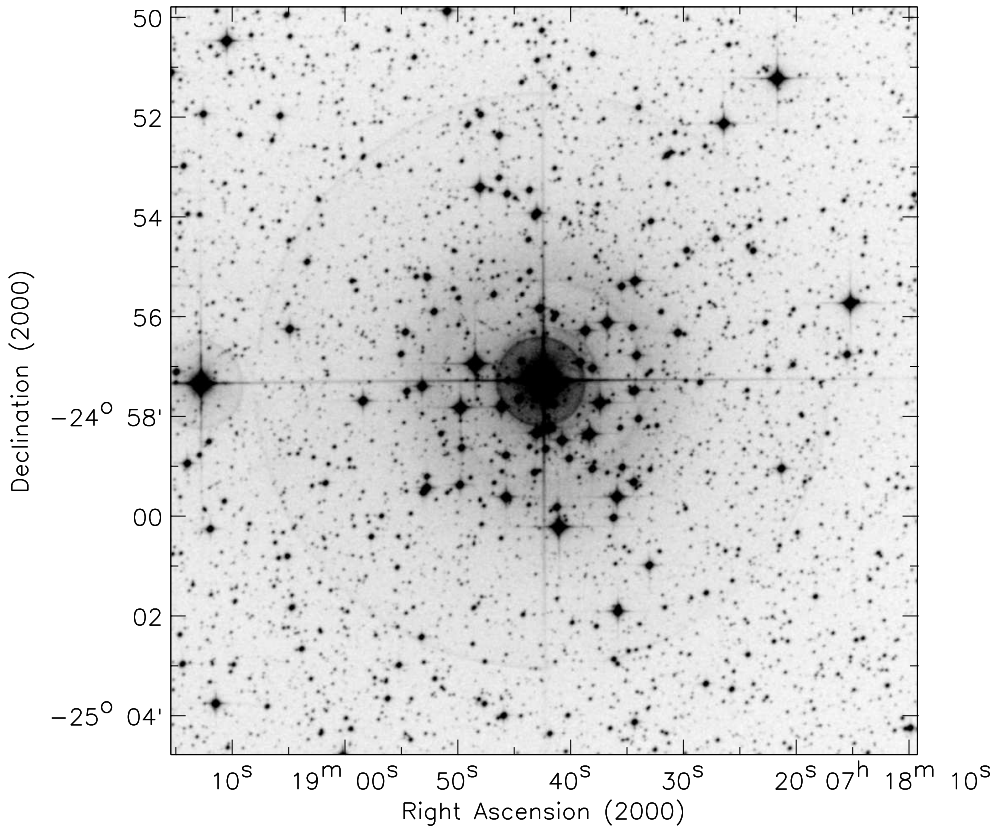}{11.6cm}{90.0}{58.0}{58.0}{70.0}{-15.0}
\caption[f1.ps]{A 15\arcmin$\times$15\arcmin\ Second Palomar
Observatory Sky Survey (POSS-II) red image of NGC\,2362 obtained from
the Digitized Sky Survey. $\tau$ CMa, the 4$^{th}$ mag O9 Ib star
(center), is the most massive cluster member and the only star that
has evolved significantly away from the ZAMS. Given the abundance of
unevolved B-type members from B1 to B9, the early population of
NGC\,2362 has often been used to define the upper section of the
empirically-derived ZAMS. The pre-main sequence population of
NGC\,2362 is symmetrically distributed around $\tau$ CMa.\label{f1}}
\end{figure}

 The young cluster NGC\,2362 in Canis Majoris (CMa) is
dominated by the 4$^{th}$ mag O9 Ib multiple star, $\tau$ CMa and
several dozen B-type stars (McSwain \& Gies 2005), spherically
distributed within a volume $\sim$3 pc in radius. Shown in Figure~1 is
a 15\arcmin$\times$15\arcmin\ Second Palomar Observatory Sky Survey
(POSS-II) red image of NGC\,2362 obtained from the Digitized Sky
Survey. The cluster is free of molecular gas and nebular emission and
suffers very little interstellar reddening despite an accepted
distance of nearly 1.5 kpc.  With an age of $\sim$5 Myr, only $\tau$
CMa has evolved significantly away from the cluster zero-age main
sequence (ZAMS). Although probably relaxed, the evaporation timescale
for the cluster is significantly greater than its age, implying that
few members have dispersed. In essence, an unobscured and
near-complete stellar population remains around $\tau$ CMa, making
NGC\,2362 an ideal target for initial mass function (IMF) studies
(Moitinho et al. 2001). The rich history of the cluster begins with
its discovery by Fr. Giovanni Battista Hodierna in Sicily during the
17$^{th}$ century using a Galilean-type refractor.  His observations
of nebulous objects which includes NGC\,2362 were published in 1654.
With only one member ($\tau$ CMa) visible to the unaided eye, the
cluster quickly returned to obscurity for over a century before Sir
William Herschel noted its presence, entry H VII.17 in his catalog of
stellar clusters and nebulae, one of the pre-cursors to Dreyer's New
General Catalog (NGC). Dreyer's notes within the original NGC
summarize NGC\,2362 as a pretty large, rich cluster centered upon 30
CMa ($\tau$). The compact nature of NGC\,2362 is quite striking when
viewed on the Palomar Observatory Sky Survey (POSS) plates. The large
number of early-type stars form a luminous halo around $\tau$ CMa,
$\sim$10\arcmin\ in diameter (Figure~\ref{f1}).

Although lacking nebulosity in the immediate cluster vicinity, just
over one degree east of NGC\,2362 extensive H~II emission is apparent
on the POSS plates. This emission is part of the giant H~II region
Sharpless 310, described by Sharpless (1959) as an incomplete ring
8$^{\circ}$ in diameter.  The ionizing sources for this gas are most
likely 29 CMa, $\tau$ CMa, and the early B-type members of
NGC\,2362. IRAS images of the NGC\,2362 region reveal that the cluster
is located within an evacuated cavity approximately 30\arcmin\ in
diameter. A partial ring of dust emission, most prominent at
60~$\mu$m, is evident to the east. The nearby dark nebula, L1667, also
believed to be 1.5~kpc distant (Lada \& Reid 1978), lies southeast of
NGC\,2362, near the variable M5 supergiant VY~CMa. Tenuous dark
clouds appear to follow the contours of H~II emission on the POSS
plates, perhaps remnants of the molecular cloud complex from which
NGC\,2362 formed. The OB stellar population of NGC\,2362 may have
triggered a second generation of star formation within L1660 to the
northeast, which hosts the Herbig-Haro object HH~72 (Reipurth \& Graham
1988).  Shown in Figure~2 is a reproduction of a Schmidt plate from
Reipurth \& Graham (1988) centered upon L1660 and with the
locations of HH~72 and three known H$\alpha$ emission stars
identified. Reipurth \& Graham (1988) conclude that L1660 is slowly
being eroded away by intense UV radiation from the OB stellar
population of NGC\,2362 and other nearby massive stars.

\begin{figure}
\centering
%\plotfiddle{f2.eps}{11.6cm}{0.0}{58.0}{58.0}{0.0}{0.0}
\includegraphics[angle=0,width=0.85\textwidth]{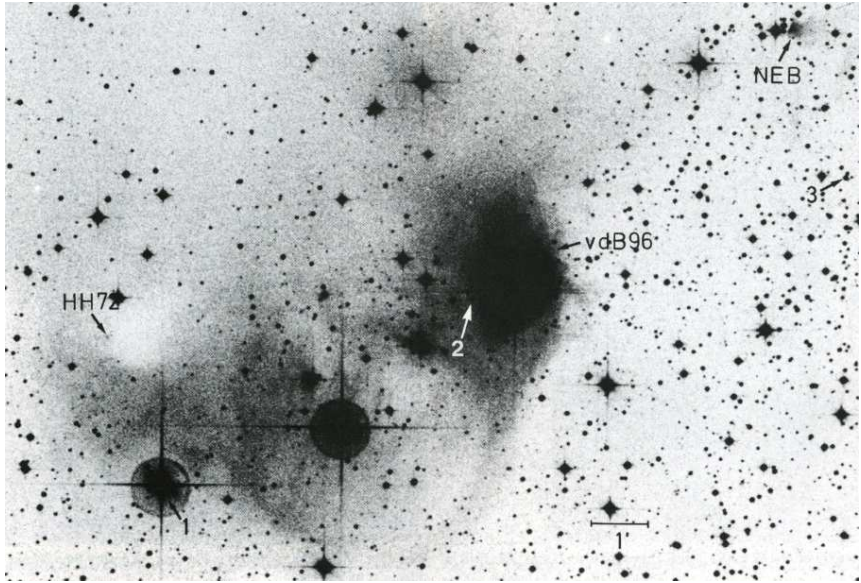}
\caption[f2.eps]{The L1660 dark cloud and HH 72 on the SERC-J
Schmidt plate from Reipurth \& Graham (1988).  The field shown is
approximately 13\arcmin$\times$17\arcmin. The locations of three
H$\alpha$ emission stars are also identified, one of which (labeled 1)
is possibly a HAeBe star. The other two H$\alpha$ emitters are
candidate low-mass T Tauri stars. \label{f2}}
\end{figure}

\section{Cluster Distance and Interstellar Extinction}

 With an abundance of early-type stars lying on or near
the ZAMS, NGC\,2362 is seemingly well-suited for precise distance
determinations. Distance estimates for the cluster, however, vary
significantly in the literature from Humphreys' (1978) value of 904~pc
(for $\tau$ CMa) to nearly 2100~pc by Johnson \& Morgan (1953). The
difficulty in fixing the cluster distance arises from the near
vertical slope of the ZAMS for the early-type cluster members in the
color-magnitude diagram (CMD). Most recently, Moitinho et al. (2001)
used CCD photometry to establish a distance modulus of 11.16 mag for
the cluster by fitting the upper main sequence with the Schmidt-Kaler
(1982) ZAMS in the $V$, $U-B$ plane. The $U-B$ color yields the
shallowest slope for the B-type stellar sequence, thereby permitting a
precise ZAMS fit.  From their derived interstellar reddening value of
$E_{B-V}$$=$0.1 mag, and by assuming the standard ratio of
total-to-selective absorption, $R_{V}$$=$3.1, Moitinho et al. (2001)
determined a distance of 1480~pc, which nearly matches that derived by
Balona \& Laney (1996).  Interstellar extinction along the line of
sight to NGC\,2362 is quite low given the assumed distance of the
cluster.  All investigations of the cluster are consistent with
$E_{B-V}\sim$0.1 mag and none exhibit evidence for variable extinction
across the face of the cluster. Numerous background galaxies are
apparent in deep optical images of the field, providing additional
circumstantial support for the lack of significant interstellar
extinction along the line of sight toward the cluster. Table~1
summarizes the distance and extinction estimates determined by
selected investigations of NGC\,2362. Also included in the table are
the adopted ages and evolutionary models used for the cluster age
determinations.

\section{The Age of NGC\,2362}

\begin{figure}[tb]
%\hspace{-1.5cm}
%\plotfiddle{f3.ps}{10.0cm}{90.0}{58.0}{58.0}{50.0}{-30.0}
\centering
\includegraphics[draft=False,angle=90,width=0.9\textwidth]{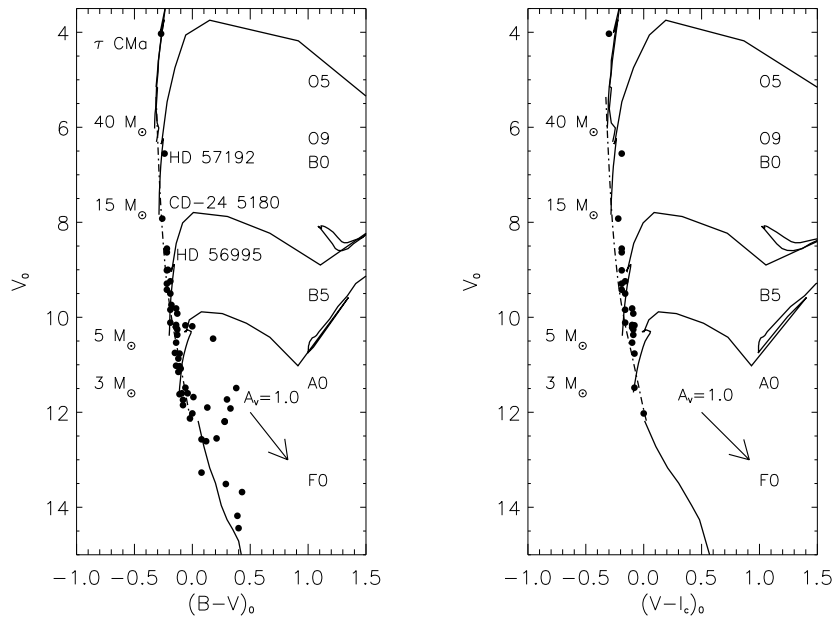}
\caption[f3.ps]{The $(B-V)_{0}$, $V_{0}$ (left) and $(V-I_{c})_{0}$,
$V_{0}$ (right) CMDs for the OB stellar population of
NGC\,2362. Photometry and spectral types for these stars were obtained
from the literature. The solid line rising from the bottom edge of
each figure is the Pleiades main sequence from Stauffer (1984). The
dashed line extending above the Pleiades fit is the main sequence
derived by Balona \& Shobbrook (1984). Peeling away from the ZAMS are
the post-main sequence evolutionary tracks of Schaller et al. (1992)
for 40, 15, 5, and 3~M$_{\odot}$ stars. The base of each of these
evolutionary tracks is labeled with its corresponding mass. On the
right side of each panel are the main sequence spectral types at the
approximate $V_{0}$ assuming a distance modulus of 10.85.\label{f3}}
\end{figure}

 From the large number of unevolved B-type stars, from
B1V to B9V, the age of NGC 2362 has always been inferred to be quite
young. Balona \& Laney (1996) assigned an age of 5~Myr to NGC\,2362 by
using the post-main sequence evolutionary models of Meynet et
al. (1993) to fit $\tau$ CMa, the only evolved cluster member. The
upper main sequence of the cluster is shown in Figure~3, the
$(B-V)_{0}$, $V_{0}$ (left panel) and the $(V-I)_{0}$, $V_{0}$ (right
panel) CMDs plotted using photometry and spectral types available in
the literature.  The sources for the photometric data include Johnson
(1950), Johnson \& Morgan (1953), Vandenbergh \& Hagen (1968), and
Blanco et al. (1968). Available spectral types are from numerous
sources including Johnson \& Morgan (1953), Hiltner, Garrison, \&
Schild (1969), and Houk \& Smith-Moore (1988). The solid line is the
Pleiades main sequence from Stauffer (1984), placed at the adopted
cluster distance of 1480~pc. The ZAMS derived by Balona \& Shobbrook
(1984) is shown as a dashed line extending above the Pleiades fit. The
ZAMS appears to fit the upper portion of the $(B-V)_{0}$, $V_{0}$ CMD
well with two exceptions, $\tau$~CMa and HD~57192. As previously
stated, $\tau$~CMa is known to be an evolved multiple star, but
HD~57192 (B2 V), assuming its published luminosity class, would be
overluminous if at the distance of NGC\,2362. Kazarovets et al. (1999)
identify the star as an eclipsing $\beta$ Lyrae-type binary, which
may account for its additional flux assuming identical masses for the
two components. Lying $\sim$7\arcmin\ east of $\tau$~CMa, it is
possible that the star is not a cluster member, however, its radial
velocity does match that of other members (Evans 1967).

Superposed in Figure~3 are the post-main sequence evolutionary tracks
of Schaller et al. (1992) for 40, 15, 5, and 3 M$_{\odot}$ stars. From
the models, the age at which a significant departure from the main
sequence first becomes detectable can be estimated for the various
stellar masses. If it is assumed that a shift in color of $\sim$0.1
mag.  would be detectable, the ages at which such evolution has
occurred are: 4.3 Myr (at 40 M$_{\odot}$), 6.4 Myr (25 M$_{\odot}$),
and 11.6 Myr (15 M$_{\odot}$). There is no evidence for evolution away
from the ZAMS among the B-type stars, the earliest of which is the B1
dwarf, CD$-$24$^\circ$5180 (Johnson's star 30). $\tau$~CMa is seen to lie on
the 40~M$_{\odot}$ track of Schaller et al. (1992), suggesting an age
of $\sim$4.1~Myr if this mass is assumed and its multiple nature
ignored.  This value agrees well with the post-main sequence age of
$\tau$~CMa derived by Moitinho et al. (2001) using the models of
Girardi et al. (2000). Assuming coevality, significant evolution
should not have occurred for any other cluster member given the

\begin{sidewaystable}[htbp]
\begin{footnotesize}
\begin{center}
%\begin{landscape}
\begin{tabular}[c]{c@{\hskip4pt}ccccc@{\hskip4pt}c}
\multicolumn{7}{c}{Table 1} \\
\multicolumn{7}{c}{Summary of Previous Investigations of NGC\,2362} \\
\hline \hline
Authors & Age (Myr) & M$_{V}$ Range & Isochrone$^{a}$ & E($B-V$ & Distance (pc) & Notes\\
\hline
Johnson (1950)           & ...     &  $-$7.0 $\le$ M$_{V}$ $\le$ +4.0  & ...          & ...  & 1410 & UBV photoelectric\\
Johnson \& Morgan (1953) & ...     &  $-$7.0 $\le$ M$_{V}$ $\le$ +4.0  & ...          & 0.10 & 2090 & UBV photometry / spectroscopy\\
Fenkart (1962)           & ...     &  ...                              & ...          & 0.08 & 1620 & ugr\\
Perry (1973)             & ...     &  $-$7.0 $\le$ M$_{V}$ $\le$ +1.0  & ...          & 0.10 & 1600 & uvby$\beta$ photoelectric\\
Mermilliod (1981)        & ...     &  $-$7.0 $\le$ M$_{V}$ $\le$ +1.5  & ...          & 0.11 & 1380 & Galactic cluster study\\
Mermilliod \& Maeder (1986) & 7.0  &  $-$7.0 $\le$ M$_{V}$ $\le$ +1.5  & Maeder (1981)& 0.11 & 1622 & UBV photoelectric\\
Balona \& Laney (1996)   & 5.0     &  $-$7.0 $\le$ M$_{V}$ $\le$ +4.5  & Meynet et al. (1993) & 0.10 & 1490 & uvby$\beta$ CCD\\
Moitinho et al. (2001)   & 5.0     &  $-$2.0 $\le$ M$_{V}$ $\le$ +10.0 & B98 \& G00   & 0.10 & 1480 & UBVRI CCD\\
Dahm (2005)              & 3.5-5.0 &  $-$2.0 $\le$ M$_{V}$ $\le$ +10.0 & DM97 \& B98  & 0.10 & 1480 & VRI CCD and spectroscopic\\
\hline \hline
\end{tabular}
\begin{flushleft} $^{a}$ B98 - Baraffe et al. (1998), DM97 - D'Antona \& Mazzitelli (1997), G00 - Girardi et al. (2000)\\
\end{flushleft}
%\end{landscape}
\end{center}
\end{footnotesize}
\end{sidewaystable}

\noindent Schaller et al. (1992) timescales. Alternatively, one could estimate
the cluster age using contraction times for the latest B-type stars
lying on the ZAMS. Inspection of Figure~3 suggests that above
$V_{0}\sim$11, or M$_{V} \sim -$0.2 (B8), all stars lie in close
proximity to the ZAMS.  The mean mass for a B8 dwarf is $\sim$3.0
M$_{\odot}$ (Andersen 1991), implying a theoretical contraction time
of 7.2~Myr (Bernasconi \& Maeder 1996). If, however, the B8 stars are
still settling onto the ZAMS, and the latest zero-age main sequence
stars were of type B5 (5.0~M$_{\odot}$), contraction times of 1.2~Myr
follow from the models of Bernasconi \& Maeder (1996). Uncertainties
in both theory and observation, however, certainly dominate.

Moitinho et al. (2001) first employed the use of pre-main sequence
isochrones to estimate the age of NGC\,2362 by fitting the narrow
sequence of low-mass stars with the models of Baraffe et
al. (1998). Comparing the post-main sequence evolutionary models of
Girardi et al. (2000) with the pre-main sequence isochrones of Baraffe et al. (1998),
Moitinho et al. (2001) found agreement with an age of 5$^{+1}$$_{-2}$
Myr. Dahm (2005) fit the narrow pre-main sequence of H$\alpha$
emitters with the models of Baraffe et al. (1998) and D'Antona \& Mazzitelli
(1997). Differences in median ages between the models were noted by
Dahm (2005), but an age of 3.5--5~Myr is in best agreement with the
post main sequence age of $\tau$~CMa and the main sequence contraction
times for mid-to-late B-type stars.  Perhaps most remarkable of the
fundamental parameters of NGC\,2362, however, is not its youth, but
rather the small age dispersion evident within the cluster. Both
Moitinho et al. (2001) and Dahm (2005) estimate an age dispersion of
less than 3~Myr among the low-mass population, suggesting that star
formation occurred rapidly, within a single burst. The coeval nature
of the pre-main sequence population extends from the substellar limit
to solar mass and intermediate mass stars. Much of the apparent
dispersion in the CMD can be accounted for by multiplicity or
observational errors, suggesting an even lower age spread. Similar
narrow sequences are observed in other Galactic clusters, but
NGC\,2362 is among the least evolved to exhibit such a well-defined,
coeval population.

\section{The OB-Stellar Population of NGC\,2362}

 As noted previously, the most massive member of
NGC\,2362 is $\tau$~CMa (HD~57061, 30~CMa). Trumpler (1935) noted a
radial velocity difference of $+$9.9~km~s$^{-1}$ between it and other
cluster members and interpreted this difference as a gravitational
redshift induced by the star's mass, calculated to be over
300~M$_{\odot}$.  More recent measurements have reduced the difference
between the star's $\gamma$ velocity and the cluster radial velocity
to $\sim$8.0~km~s$^{-1}$ (Van Leeuwen \& van Genderen 1997). Some
uncertainty can be accounted for by the multiple nature of $\tau$
CMa. Struve \& Pogo (1928) and Struve \& Kraft (1954) found $\tau$
CMa to be a single-line, spectroscopic binary (SB1), while Finsen
(1952) discovered $\tau$~CMa to be a visual pair composed of two stars
separated by 0\farcs158. The Hipparcos catalog gives a separation of
0\farcs152 for the two stars and Hipparcos passband magnitudes of
4.887 and 5.329. Van Leeuwen \& van Genderen (1997) propose that
$\tau$ CMa is a quadruple system, a composite of two binaries: the
visual pair of O stars, the SB1 secondary of the brighter of the
visual pair with a 154.9 day period and an eclipsing component with a
period of 1.28 days. They further suggest that the composite system
was formed through a merger of existing binaries. The multiple nature
of $\tau$ CMa and its evolved state certainly complicates
interpretation of its position on the CMD.

The B-type main sequence population of NGC\,2362 is believed to number
around 40 stars, but membership cannot be confirmed without spectral
type information or radial velocities for many of the candidate
members.  Surprisingly little modern work has been done on the B-star
population of the cluster, with many of the available spectral types
dating back to Johnson \& Morgan (1953). Perry (1972) lists 14
early-type stars in the region as non-members on the basis of $UBV$
photometry, but only nine of these lie within 5\arcmin\ ($\sim$2 pc)
of the cluster core. Of these nine, at least two are visual pairs and
two others lie within 1$\arcmin$ of $\tau$ CMa, raising the
possibility of photometric errors induced by scattered light. Balona
\& Laney (1996) presented $uvby\beta$ CCD photometry of the cluster
upper main sequence in their attempt to identify short-period $\beta$
Cep-type variables. Although none were found, the photometry revealed
a narrow ZAMS consisting of at least 25 stars extending from
$V\sim$7--12. The most recent investigation of the early stellar
population of NGC\,2362 is that of McSwain \& Gies (2005) who examined
41 OB-type stars within the cluster vicinity for possible H$\alpha$
emission using

\noindent Str\"omgren photometry. Only one candidate member,
CD$-$24$^\circ$5166, was identified as a possible emission source. Given the
young age of NGC\,2362, the dearth of classical Be stars within the
cluster is consistent with the Be frequency-age dependence. NGC\,2362
may lie within a narrow age region when the Herbig AeBe phenomenon has
subsided and the classical Be line emission has not yet appeared.
Table~2 lists approximately 40 early-type stars in the vicinity of
$\tau$~CMa with their J2000 coordinates, spectral types, and $B-$ and
$V-$band photometry from SIMBAD. Spectral types for some stars taken
from McSwain \& Gies (2005) are listed simply as ``B,'' and should be
regarded as uncertain.

\begin{table}[htbp]
\begin{center}
\begin{tabular}[c]{cccccc@{\hskip2pt}c}
\multicolumn{7}{c}{Table 2} \\
\multicolumn{7}{c}{Candidate Early-type Members of NGC\,2362} \\
\hline \hline
Nr.$^{a}$ & RA (J2000) & $\delta$ (J2000) & SpT & $B$ & $V$ & Other Identifiers\\
\hline
68      &       07 18 21.94 & --24 51 11.9 & B3 IV/V & 8.80 & 8.94 & HD 56995\\
78      &       07 19 26.20 & --24 56 31.8 & B:    & 10.73 & 10.71 & CD$-$24 5192\\
69      &       07 18 26.66 & --24 52 06.4 & Be?  & 9.99  & 10.06 & CD$-$24 5166\\
1       &       07 18 33.06 & --25 00 57.1 & B:    & 11.39 & 11.40 & CPD$-$24 2195\\
2       &       07 18 34.26 & --24 56 45.6 & B:    & 11.16 & 11.17 & CPD$-$24 2198\\
5       &       07 18 34.45 & --24 55 15.7 & B5 V  & 10.72 &  10.76 & CPD$-$24 2196\\
66      &       07 18 34.48 & --24 59 17.1 & B:    & 11.44 & 11.46 & CPD$-$24 2199\\
11      &       07 18 35.62 & --24 55 22.3 & B:    & 11.95 & 11.93 & CPD$-$24 2200\\
13      &       07 18 35.82 & --25 01 52.6 & B:   & 10.59 & 10.59 & CPD$-$24 2201\\
9       &       07 18 35.95 & --24 59 35.0 & B3 V  & 9.76  &  9.80 & CD$-$24 5170\\
12      &       07 18 36.89 & --24 56 05.8 & B5 V  & 9.99  &  10.05 & CD$-$24 5172\\
48      &       07 18 37.48 & --24 57 42.2 & B3 V  & 9.43  &  9.54 & CPD$-$24 2205\\
15      &       07 18 38.13 & --24 59 01.5 & B9 V  & 11.79 &  11.75 & CD$-$24 5171 \\
14      &       07 18 38.41 & --24 58 20.0 & B2 V  & 9.43  &  9.6 & CPD$-$24 2207 \\
16      &       07 18 38.81 & --24 56 15.6 & B9 V  & 10.64 &  10.57 & QY CMa   \\
70      &       07 18 40.19 & --24 58 49.3 & B:    & 11.92 & 11.90 & CPD$-$24 2211 \\
21      &       07 18 40.83 & --24 58 27.4 & B6 V  & 10.37 &  10.43 & CPD$-$24 2212    \\
20      &       07 18 41.07 & --25 00 11.4 & B2 V  & 8.58  &  8.77 & CD$-$24 5175     \\
95      &       07 18 41.55 & --24 57 44.8 & B:    & ...   & 11.53 & ...     \\
39      &       07 18 41.99 & --24 58 12.3 & B2 V  & 9.69  &  9.78 & ...     \\
22      &       07 18 42.26 & --24 58 38.0 & B:    & 11.95 & 11.91 &  ...    \\
23      &       07 18 42.49 & --24 57 15.8 & O9 Ib & 4.25  &  4.39 & $\tau$ CMa\\
24      &       07 18 42.86 & --24 55 49.1 & B7 V  & 10.97 &  10.98 & CPD$-$24 2217  \\
50      &       07 18 43.12 & --24 58 18.9 & B6 V  & 10.20 &  10.20 & CPD$-$24 2220      \\
25      &       07 18 43.14 & --24 53 54.9 & B5 V  & 10.78 &  10.77 & CPD$-$24 2218 \\
26      &       07 18 45.77 & --24 59 35.7 & B7   & 10.36 &  10.38 & CPD$-$24 2223 \\
52      &       07 18 45.78 & --24 58 45.5 & B:    & 11.97 & 11.91 & CPD$-$24 2222   \\
27	&	07 18 46.20 & --24 57 47.6 & B3 V	 & 10.10 &  10.15 & CPD$-$24 2225	\\
30      &       07 18 48.54 & --24 56 56.0 & B1 V  & 8.04  &  8.21 &  CD$-$24 5180     \\
28      &       07 18 49.70 & --24 58 36.8 & B:    & 12.07 & 12.05 & CPD$-$24 2229     \\
31	&	07 18 49.83 & --24 57 48.7 & B2 V	 & 9.20  &  9.31 & IM CMa	\\
29      &       07 18 49.84 & --24 59 21.3 & B:    & 11.31 & 11.33 & CPD$-$24 2230  \\
56      &       07 18 52.79 & --24 55 12.2 & B:    & 12.14 & 12.01 & CPD$-$24 2234  \\
34	&	07 18 53.22 & --24 57 23.2 & B5 V	 & 10.47 &  10.50 & ...	\\
58      &       07 18 54.57 & --24 57 29.1 & A0   & 12.30 & 12.23 & ...     \\
57      &       07 18 54.71 & --24 56 18.1 & B:    & 12.18 & 12.16 & CPD$-$24 2237     \\
36	&	07 18 58.44 & --24 57 41.1 & B3 V	 & 10.78 &  10.76 & CPD$-$24 2240	\\
42      &       07 19 04.93 & --24 56 15.1 & B:    & 11.32 & 11.34 & CPD$-$24 2244      \\
Anon    &       07 19 12.62 & --24 51 57.0 & B:    & ...   & 12.17 & ... \\
46	&	07 19 12.77 & --24 57 20.5 & B2 V& 6.630 &  6.801 &  HD 57192\\
76      &       07 19 16.75 & --24 53 31.4 & B:    & 9.76  & 9.80  & CPD$-$24 2250     \\
Anon    &       07 19 19.39 & --24 52 29.4 & B:	 & ...	 & 11.95 & ...	\\
\hline \hline
\multicolumn{7}{l}{\parbox{0.9\textwidth}{\footnotesize $^a$
 Number from Johnson (1950) }}\\[2ex]

\end{tabular}
\end{center}
\begin{flushleft} \end{flushleft}
\end{table}

\section{The Initial Mass Function}

 Interest in NGC\,2362 was renewed by Baade who remarked
to Johnson (1950) that the cluster is composed almost exclusively of
early-type stars and devoid of nebulosity. This point was emphasized
by Baade in a lecture series given at Harvard Observatory in 1958 in
which he used NGC\,2362 as an example of a cluster with a mass
function that clearly deviates from that of the field (Baade
1963). Consequently, Johnson (1950) conducted the first modern
investigation of the cluster, determining a distance modulus from
photoelectric magnitudes of the brightest cluster members of $m_{V} -
M_{V} = 10.75$ (1410~pc). In their study of Galactic cluster
luminosity functions, Vandenbergh \& Sher (1960) supported Baade's
statement regarding the dearth of intrinsically low-luminosity stars
in NGC\,2362, but conceded that the photographic plates of the cluster
obtained at Palomar Observatory were affected by the lights of San
Diego. Another limitation discussed by Vandenbergh \& Sher (1960) was
their inability to consistently reproduce star counts, a systematic
error that certainly scaled with stellar magnitude.  In retrospect,
the low response and poor quality of the photographic plates, the
overwhelming brightness of the early-type cluster members, and the
human factor introduced by the star counting process probably all
contributed to the non-detection of the cluster's faint stellar
population. Consequently, Baade's early commentary and the conclusions
of Vandenbergh \& Sher (1960) remained undisputed for another three
decades.

The advent of CCD detectors brought about a revolution in
observational astronomy, and with it a renewed search for an
intrinsically faint stellar population around $\tau$ CMa. In March
1990 Wilner \& Lada (1991) imaged NGC\,2362 with the 90 in. telescope
of Steward Observatory using an 800$\times$800 pixel CCD camera.  The
challenges involved in direct imaging of the cluster, however, are
considerable and deserve brief mention. Centrally located, the
4$^{th}$ mag $\tau$~CMa causes severe scattered light and charge
blooming problems even on the shortest of integrations. These
difficulties can be overcome with various imaging strategies
(quadrants around $\tau$ CMa, combining multiple short exposures,
etc.), but the faint stars in close proximity of $\tau$ CMa remain
extremely difficult to observe. Wilner \& Lada (1991) examined
NGC\,2362 using deep $I-$band imaging complete to
m$_{I}$$\sim$16.6. The interior 1.6$'$ of the cluster were excluded from
the study because of severe internal reflections, but for the first
time, a substantial population of faint stars (415) was detected
around $\tau$ CMa. Despite the presence of these stars, however,
Wilner \& Lada (1991) concluded that a deficit of low-mass ($<$0.8
M$_{\odot}$) stars relative to the Salpeter (1955) IMF was
apparent. One possible explanation put forth by Wilner \& Lada (1991)
was that if the duration of star formation within NGC\,2362 was longer
than the age inferred from the evolutionary state of the B-type stars,
mass segregation may have occurred, effectively moving low-mass
members to the outer cluster regions where the survey was
incomplete. Kroupa, Gilmore, \& Tout (1992) re-examined the luminosity
function (LF) of NGC\,2362 presented by Wilner \& Lada (1991),
comparing it to the solar neighborhood mass function regressed to an
age of 10~Myr. Their modeled LF was found to be in agreement (within
error) with that observed for NGC\,2362, suggesting that the cluster
mass function is similar to that of the field star population. The
full extent of the low-mass population was not realized until Moitinho
et al. (2001) presented their CMD of the cluster. Recently, however,
the possibility of a non-standard IMF for NGC\,2362 was again raised
by the deep {\it Chandra} survey of the cluster by Damiani et
al. (2006b). Their analysis of cluster X-ray sources finds a real
deficit of low-mass stars compared to a power law or a log-normal
distribution. When compared to IC\,348 or the Orion Nebula Cluster,
the IMF of NGC\,2362 appears to be significantly different.

\begin{figure}[!htbp]
\centering
\includegraphics[draft=False,angle=0,height=0.5\textheight]{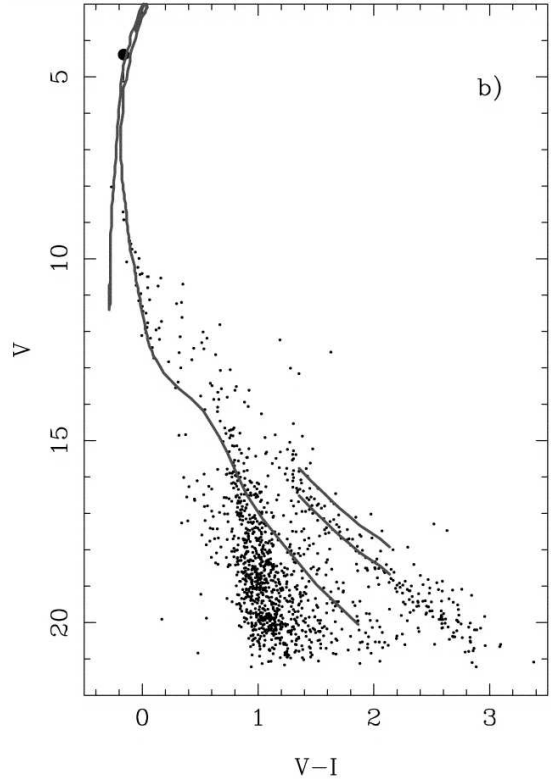}
\caption[f4.ps]{ A $V$, $V-I$ color-magnitude diagram of NGC\,2362,
from Moitinho et al. (2001). The line on the left is the 5 Myr
post-main sequence isochrone from Girardi et al. (2000) and the lines
on the right are the 5 Myr pre-main sequence isochrone of Baraffe et
al. (1998) and the upper limit of the binary sequence, 0.75 mag. more
luminous.  The large solid point in the top-left corner of the figure
represents $\tau$ CMa. \label{f4}}
\end{figure}

\section{The Low-Mass Population of NGC\,2362}

 The $UBVRI$ photometric survey of NGC\,2362 of Moitinho
et al. (2001) revealed the long, narrow pre-main sequence of stars
extending more than 9~mag in the CMD from $V\sim11$ to 20,
corresponding to early-A spectral types to nearly the hydrogen burning
limit. Shown in Figure~4 is the $V$, $V-I$ CMD of NGC\,2362 taken from
Moitinho et al. (2001). The solid line to the left is the 5~Myr
post-main sequence isochrone of Girardi et al. (2000), and the two
parallel lines to the right are the 5~Myr isochrone of Baraffe et
al. (1998) and its complementary binary sequence limit lying
0.75~mag. above. The Baraffe et al. (1998) 5~Myr isochrone fits the
cluster pre-main sequence well, but is truncated near $V-I$$\sim$2.2
to avoid complications with the models. The deep $VRI$ photometric
survey by Dahm (2005) extended the cluster pre-main sequence another
two magnitudes, finding nearly 500 stars (undoubtedly including many
field interlopers) lying above the ZAMS. Shown in the left panel of
Figure~5 is the $V$, $V-I$ CMD for all stars in the Dahm (2005)
photometric survey. The narrow pre-main sequence discovered by
Moitinho et al. (2001) is strongly evident and densely populated, to
the substellar limit. The solid line in the panel is the Pleiades main
sequence and the upper main sequence of Balona \& Shobbrook (1984).

\begin{figure}[!hbp]
\centering
\includegraphics[draft=False,angle=90,width=\textwidth]{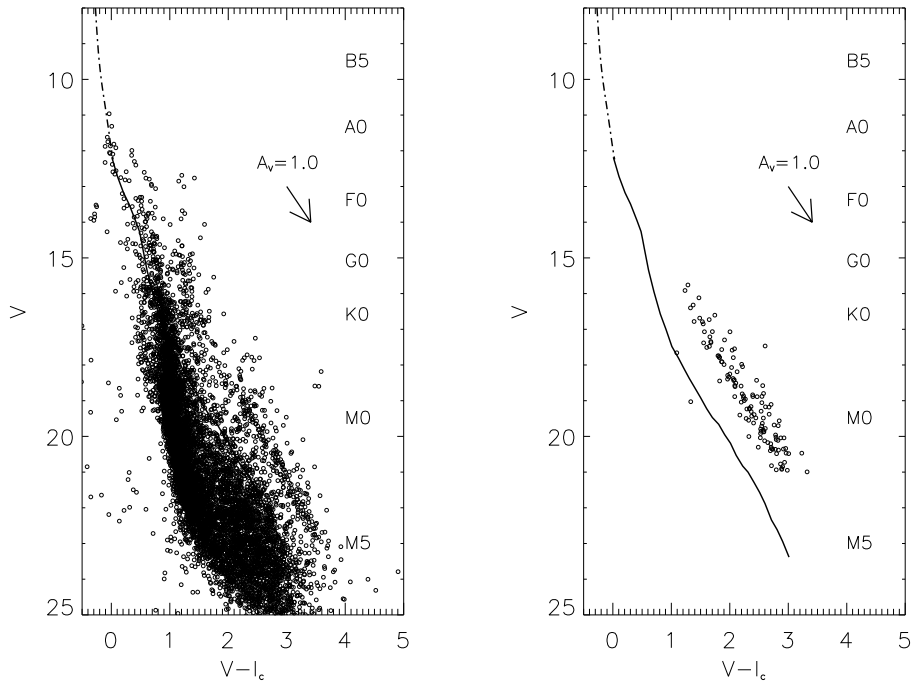}
\caption[f5.ps]{The $V-I_{c}$, $V$ color-magnitude diagrams of
NGC\,2362 for all stars observed by Dahm (left) and for just the
H$\alpha$ emitters (right). No allowance has been made for
interstellar extinction. The solid line in each panel is the Pleiades
main sequence from Stauffer (1984) and placed at the adopted cluster
distance of 1480~pc. \label{f5}}
\end{figure}

Given the relative youth of the cluster, the low-mass stellar
population of NGC\,2362 should still be in its T Tauri phase of
evolution, either actively accreting circumstellar gas or experiencing
enhanced chromospheric activity. G. H. Herbig (private communication)
first undertook a slitless grism survey of the cluster in 1991, and
despite the low dispersion of the survey, 6.6\AA\ pixel$^{-1}$, two
H$\alpha$ emitters were detected in the cluster.  Follow-up slitless grism and
Gemini multi-object spectrograph (GMOS) surveys by Dahm (2005)
detected an additional $\sim$130 H$\alpha$ emitters. These
moderate-resolution GMOS spectra were also adequate for the
determination of Li~I $\lambda$6708 line strengths, thereby permitting
confirmation of youth for many of the H$\alpha$ emission
population. Most of the H$\alpha$ emitters identified within the
cluster lie along the narrow pre-main sequence of stars in the CMD, as
shown in the right panel of Figure~5. Comparing the left and right
panels of Figure~5, we see that perhaps 200--300 more low-mass
candidates lie near the established cluster sequence. Deeper optical
surveys are currently underway by several groups, which will almost
certainly identify most low-mass and very low-mass cluster members.
The Monitor project of Irwin et al. (2008) is a deep, time-series
photometric survey of NGC\,2362 used to derive rotation periods
for 271 cluster members with masses between 0.1 and 1.2 M$_{\odot}$.
The same data set was incorporated into the search for transiting
planets by Miller et al. (2008) which identified six stars with potential
planetary transit events.

\section{The Cluster Mass}

 Given the apparent absence of molecular gas and H II
emission in the cluster vicinity, the mass of NGC\,2362 should be
dominated by its stellar population. The first attempt at establishing
a cluster mass was made by Lohmann (1977) who performed a star count
and LF analysis of the cluster and derived a total cluster mass of 246
M$_{\odot}$ by summing down to $M_{V}$$\sim$5.8. Bruch \& Sanders
(1983) subsequently used Lohmann's mass for NGC\,2362 as a calibrator
in their determination of absolute masses of open clusters and OB
associations.  Muench (2002) constructed the $K-$band luminosity
functions of NGC\,2362, IC\,348, and the Orion Nebula Cluster,
finding the underlying mass functions of each to be remarkably
similar. While the total stellar mass of NGC\,2362 is certainly less
than that of the ONC (930-1860 M$_{\odot}$: Hillenbrand 1997), it is
probably somewhat greater than that of IC\, 348 ($\sim$200
M$_{\odot}$: Lada \& Lada 1995). Mass estimates for the quadruple
system that comprises $\tau$ CMa alone range from 30 to 90 M$_{\odot}$
(van Leeuwen \& van Genderen 1997). Assuming that only early-type
stars on or near the ZAMS of the cluster are bona fide members, a total
of $\sim$200 M$_{\odot}$ can be accounted for within the cluster's
B-star population. When added to the mass of $\tau$~CMa, this value
agrees well with the cluster mass derived by Lohmann (1977).

To estimate the total mass of the cluster pre-main sequence
population, the masses for the individual H$\alpha$ emitters were
summed by Dahm (2005) to yield a model-dependent value of 45--72
M$_{\odot}$. This is certainly a lower-limit given the larger
population suggested by the CMD. An upper limit for the total mass of
the pre-main sequence population is given by the sum of masses for all
stars lying between the 1 and 10~Myr isochrones of Baraffe et al. (1998). This estimate
very likely includes many field interlopers and excludes some stars
that appear overluminous due to multiplicity. The total mass of the
$\sim$500 stars falling within these isochrones is $\sim$300
M$_{\odot}$. Adding the derived upper and lower limits for the
pre-main sequence population to the mass of the B-stars and the mass
of $\tau$~CMa, the total cluster mass probably ranges from $\sim$300
to 540~M$_{\odot}$. These estimates, however, do not account for the A
and early F-type stars, which are still on their radiative
tracks. Reviewing the CMDs shown in Figures 4 and 5, around two dozen
stars have been excluded from the pre-main sequence analysis, which
may be members of this intermediate-mass population. Assuming an
average mass of 2 M$_{\odot}$ for each, this would amount to another
$\sim$50 M$_{\odot}$. Allowing for multiplicity, outlying members, and
the very low-mass population below the completeness limit of the
photometric surveys, the stellar mass of the cluster is likely
substantially greater.

\section{Infrared Observations of NGC\,2362}

\begin{figure}[b!]
\hspace{1.0cm}
\includegraphics[angle=90,width=4in]{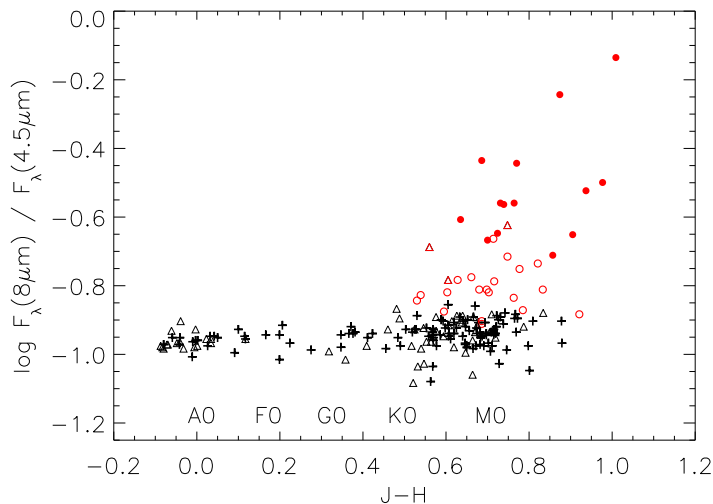}
\caption[f6.ps]{From Dahm \& Hillenbrand (2007), the IRAC-derived 8.0
to 4.5~$\mu$m flux ratio plotted as a function of $J-H$ color for all
suspected members of NGC\,2362 with 8.0 and 4.5~$\mu$m photometric
errors of $<$0.2 mag. The larger scatter about the abscissa for the
lower mass stars arises from sensitivity limits at 8~$\mu$m, which
tend to deflect the distribution toward more negative flux ratios.
None of the higher mass stars with $J-H <$ 0.5 (K2 spectral type)
exhibit significant infrared excess for $\lambda$ $<$ 8.0~$\mu$m.
\label{f6}}
\end{figure}

 Haisch, Lada, \& Lada (2001) included NGC\,2362 in
their $JHKL-$band survey of young clusters and concluded that the disk
fraction of low-mass members as inferred from infrared excesses is 12
$\pm$ 4\%. From the ages of their cluster sample, Haisch et al. (2001)
claimed an upper limit for inner disk lifetimes of $\sim$6~Myr, with
half of all stars losing their disks within 3~Myr. Their findings were
critically dependent upon the disk fraction of NGC\,2362, where their
$L-$band survey was complete only to $\sim$1~$M_{\odot}$. These
ground-based observations, however, are sensitive to only the
innermost disk regions ($\ll$1 AU). The infrared excess-derived inner
disk frequency does agree well with the fraction of TTSs within the
cluster that are classical TTSs, $\sim$9\%. While some of these strong
H$\alpha$ emitters can be accounted for by enhanced chromospheric
activity, a handful are definitive accretors as evidenced by their
complex optical spectra. {\it Spitzer} InfraRed Array Camera (IRAC)
observations of NGC\,2362 were used by Dahm \& Hillenbrand (2007) to
examine 232 suspected cluster members drawn from known H$\alpha$
emission stars, X-ray-detected stars from the 100~ks archival {\it
Chandra} observation, and established lithium-rich stars to identify
the remnant disk-bearing population of the cluster.  Dahm \&
Hillenbrand (2007) derive an upper limit for the primordial, optically
thick disk fraction of NGC\,2362 of $\sim 7\pm 2$\%, with another
$\sim 12\pm 3$\% of suspected members exhibiting infrared excesses
indicative of weak or optically thin disk emission. The presence of
circumstellar disks among candidate members of NGC\,2362 was also
found to be strongly mass-dependent, such that no stars more massive
than $\sim$1.2M$_{\odot}$ exhibit significant infrared excess
shortward of 8 $\mu$m.  This is clearly demonstrated in Figure 6,
which plots the logarithm of the ratio of the 8.0 and 4.5 $\mu$m
fluxes as a function of $J-H$ color, a tracer of stellar photospheric
emission.  From H$\alpha$ emission line strengths, Dahm \& Hillenbrand
(2007) placed an upper limit for the accretion fraction of the cluster
at $\sim$5\%, with most suspected accretors being associated with
primordial, optically thick disks.  The low-mass population of
NGC\,2362 is passing through a critical era in the empirically
established disk evolutionary scenario when most stars have already
dissipated their inner circumstellar disks. NGC\,2362 has yet a vital
role to play in our understanding of disk dissipation timescales and
the formation of planetary systems.

\section{X-ray Observations of NGC\,2362}

 X-ray observations of NGC\,2362 were first made by
Berghofer \& Schmitt (1998) using ROSAT to examine the low-mass
population of the cluster. In an 83~ks PSPC integration, 229 sources
were detected within a radius of 36\arcmin\ of $\tau$ CMa. The inner
2\arcmin\ remained unresolved, but a later 87~ks exposure with HRI was
able to isolate sources within close proximity of $\tau$ CMa. The most
luminous X-ray sources of the ROSAT survey were the two O-stars,
$\tau$~CMa and 29~CMa lying $\sim$30\arcmin\ to the north.
Correlating the X-ray detections with the Digitized Sky Survey,
Berghofer \& Schmitt (1998) concluded that the majority of emitters
were associated with optically faint ($V>$12 mag.) stars. The X-ray
luminosity function for the low-mass stars in NGC\,2362 was found to
be consistent with that found in the Chamaeleon star forming region by
Alcala et al. (1997). These early ROSAT surveys, however, were
hampered by low sensitivity and spatial resolution, making source
identification within the densely clustered environment of NGC\,2362
difficult. The {\it Chandra} X-ray Observatory has revolutionized
X-ray astrophysics with sensitivities an order of magnitude greater
than older missions, sub-arcsecond-scale spatial resolutions, and
energy-band coverage from 0.1--10.0 keV. A deep (97.9~ks) {\it
Chandra} ACIS-I observation of NGC\,2362 was completed on 2004
December 23--24 by Damiani et al. (2006a,b) who identified 387 X-ray
sources down to $log L_{X} = 29.0$ within the ACIS-I field of view.
Damiani et al. (2006b) find a significantly wider spatial distribution
of low-mass stars relative to more massive stars within the cluster,
suggesting that mass segregation is occurring. As discussed in Sect.~5,
they also conclude that the cluster mass function flattens
significantly with respect to other young clusters.

Delgado et al. (2006) using the same {\it Chandra} ACIS-I observation
in conjunction with deep $UBVR_{C}I_{C}$ and 2MASS $JHK_{S}$
photometry assign a membership status to nearly 200 X-ray detected
sources within the cluster.  Their findings suggest clearly distinct
X-ray activity behaviors between pre-main sequence and main sequence
cluster members. Among pre-main sequence stars, $L_{X}$ and $L_{Bol}$
are strongly correlated as would be expected, but main sequence
cluster members show no correlation between these two
properties. Damiani et al. (2006b) also find X-ray spectral
differences between stars brighter or fainter than $log L_{X} \sim
30.3$ such that more X-ray luminous stars exhibit a hotter (kT $\sim$
2~keV) temperature component, not present in the fainter population.
With no additional {\it Chandra} or {\it XMM-Newton} observations of
NGC\,2362 planned in the near future, it is likely that no further
enlargement of the X-ray membership sample will occur below the
current X-ray completeness limit of $\sim$0.4 M$_{\odot}$.

\section{Concluding Remarks}

 Over the last half-century, NGC\,2362 has played an
extraordinary role in our understanding of the star formation
process. Early perceptions of an anomalous mass function were
overturned with the advent of modern detectors. Deep CCD surveys of
the cluster by Wilner \& Lada (1991) and Moitinho et al. (2001)
revealed a well-populated, pre-main sequence extending more than 9~mag
to nearly the substellar limit.  There are new indications, however,
that a deficit of low-mass stars is present within the cluster,
implying that the IMF issue has yet to be fully resolved.  Recent {\it
Chandra} X-ray observations of NGC\,2362 have added several hundred
more pre-main sequence candidates to the $\sim$100$+$ suspected
members exhibiting H$\alpha$ emission or strong Li~I $\lambda$6708
absorption. Deeper optical and infrared surveys of the cluster will
also push the source detection threshold into the brown dwarf regime,
permitting a closer examination of the cluster IMF. Interest in
NGC\,2362 has also shifted to the remaining optically thick
circumstellar disks around low-mass cluster members. Near infrared and
H$\alpha$ emission surveys suggest that the inner disk regions have
dissipated for most ($\sim$90\%) of the suspected cluster members as
evidenced by the decay of near infrared excess and strong H$\alpha$
emission. {\it Spitzer} observations are beginning to resolve the
remaining questions of disk frequency within the cluster. It is
perhaps somewhat ironic that in the process of identifying and
characterizing the low-mass population of NGC\,2362 over the last
decade, the OB stars have been somewhat neglected. McSwain \& Gies'
(2005) recent Str\"omgren photometric survey of 41 OB-type stars in the
cluster region found only one candidate classical Be star. A more
thorough spectroscopic analysis of the B-star population is needed to
confirm spectral types, evaluate membership, and to examine questions
of binarity, critical to understanding the placement of stars on the
ZAMS. NGC\,2362 will likely remain a favored target for ground-based
and space-based observations. Its large, statistically-significant
population of low-mass, pre-main sequence stars, its well-defined
upper main sequence, compact structure, and lack of circumstellar and
interstellar gas and dust relative to similarly aged clusters all
contribute to the cluster's unique nature.

\vspace{0.5cm}

{\bf Acknowledgments.}  I wish to thank the referee for this paper,
Francesco Damiani, and the editor, Bo Reipurth, for many helpful
comments that significantly improved the manuscript.  SED is supported
by an NSF Astronomy and Astrophysics Postdoctoral Fellowship under
award AST-0502381.  The Digitized Sky Surveys were produced at the
Space Telescope Science Institute under U.S. Government grant NAG
W-2166. The images of these surveys are based on photographic data
obtained using the Oschin Schmidt Telescope on Palomar Mountain and
the UK Schmidt Telescope.  The plates were processed into the present
compressed digital form with the permission of these institutions.
The Second Palomar Observatory Sky Survey (POSS-II) was made by the
California Institute of Technology with funds from the National
Science Foundation, the National Geographic Society, the Sloan
Foundation, the Samuel Oschin Foundation, and the Eastman Kodak
Corporation.

\end{document}